\documentstyle[prl,preprint,aps]{revtex}

\begin{document}

\draft

\title{
Pairing and Excitation Spectrum in doped $t$-$J$ Ladders
}

\author{
Hirokazu Tsunetsugu$^{1,2}$,
Matthias Troyer$^{1,2,3}$,
and T. M. Rice$^{1}$
}

\address{
$^{1}$Theoretische Physik, Eidgen\"ossische Technische Hochschule,
8093 Z\"urich, Switzerland \\
$^{2}$Interdisziplin\"ares Projektzentrum f\"ur Supercomputing,
ETH-Zentrum, 8092 Z\"urich, Switzerland \\
$^{3}$Centro Svizzero di Calcolo Scientifico, 6924 Manno,
Switzerland
}

\date{Received}
\maketitle


\begin{abstract}
Exact diagonalization studies for a doped $t$-$J$ ladder
(or double chain) show hole pairing in the ground state.
The excitation spectrum separates into a limited number
of quasiparticles which carry charge $+|e|$ and spin ${1 \over 2}$
and a triplet mode.  At half-filling the former vanish but the latter
evolves continuously into the triplet band of the spin liquid.
At low doping the quasiparticles form a dilute Fermi gas with
a strong attraction but simultaneously the Fermi wavevector,
as would be measured in photoemission, is large.
\end{abstract}


%
%

The properties of strongly correlated electrons confined to a
ladder (or double chain) and described by $t$-$J$ or Hubbard models
have been the subject of intensive investigation recently
\cite{ED,RGS,MS,RMN,SG,DVK}. The reason lies in the unusual spin
liquid nature of the undoped parent system \cite{ED,RH,SPS}.
The key question is the evolution of the finite gap in the spin
excitation spectrum upon doping. The spin gap remains in other
spin liquids systems and is a sign of strong superconducting
fluctuations \cite{MO,MI}. A recent analysis of the $t$-$J$ ladder
using a mean field (MF) theory with Gutzwiller renormalization of
the matrix elements to account for the strong correlations,
gave a continuous evolution of the spin gap with doping \cite{MS}.
The short range resonance valence bond (RVB) state evolves into
a superconductor with modified $d$-wave symmetry within this
MF theory for the $t$-$J$ ladder. A tendency towards modified
$d$-wave superconductivity was also found in a renormalization
group calculation \cite{DVK} and in a very recent numerical study
of the Hubbard ladder although no actual enhancement of the
superconducting correlations was reported \cite{RMN}.  In this
letter we report on the properties of finite $t$-$J$ ladders up
to a size of $10 \times 2$ sites using a Lanczos diagonalization
method.  We find results which give clear evidence of hole pairing
and a modified $d$-wave RVB state in lightly doped systems in
agreement with the MF theory.  An interesting difference however
is the discontinuous evolution of the excitation spectrum upon
doping.  Upon doping new quasiparticle (QP) excitations appear
carrying both charge and spin. These excitations are in addition
to a band of spin triplets which evolve continuously away from the
undoped spin liquid. This separation of the excitation spectrum into
bound holon-spinon QP's and collective triplet excitation contrasts
with the full spin-charge separation found in a Luttinger liquid.

Another reason for especial interest is the possibility of
realizing a lattice of weakly coupled ladders in the compounds
${Sr Cu_2 O_3}$ \cite{MT} and also ${(V O)_2 P_2 O_7}$ \cite{DCJ}.
In the first case the ladders are isotropic with equal hopping
matrix elements and exchange couplings along rungs and legs and
we concentrate on this case.

The $t$-$J$ ladder we will study reads
\begin{eqnarray}
  {\cal H} =
   &&-t \sum_{j, \sigma}
        \{ [ \sum_{a=1}^2
                  c_{a \sigma}^\dagger (j)
                  c_{a \sigma} (\mbox{$j$+1})
            + c_{1 \sigma}^\dagger (j) c_{2 \sigma} (j)
           ]
            + {\rm H. c. }
        \} \nonumber\\
   &&+J \sum_{j, a}
        [ {\bf S}_{a} (j) \cdot {\bf S}_{a} (\mbox{$j$+1})
           - \textstyle{1 \over 4} n_{a} (j) n_{a} (\mbox{$j$+1})
        ] \nonumber\\
   &&+J' \sum_{j}
        [ {\bf S}_{1} (j) \cdot {\bf S}_{2} (j)
          - \textstyle{1 \over 4} n_{1} (j) n_{2} (j)
        ],
\label{eqham}
\end{eqnarray}
where $j$ runs over $L$ rungs, and $\sigma$ $(=\uparrow,\downarrow)$
and $a$ $(=1,2)$ are spin and leg indices. We take $t=1$ as the
energy units. The first term is the kinetic energy and the $J$ $(J')$
are exchange couplings along the ladder (rungs).  The local
constraint excludes double occupancy at every site
$(n_{a \uparrow} (j) n_{a \downarrow} (j) = 0)$.
Although the isotropic case $J'=J$ is of most interest, we study
also the limit of $J' \gg J, 1$, which can be easily understood
and is a good starting point to trace back to the isotropic case.
Periodic or antiperiodic boundary conditions (PBC, APBC) are used
along the ladder and then the wavevector ${\bf k}=(k_x,k_y)$
is well defined, $k_x$ $(k_y)$ are along the ladder (rungs).

To begin we summarize the properties of spin excitations at
half-filling.  In this case (\ref{eqham}) reduces to the Heisenberg
spin ladder (aside from a constant). Several different approaches
have shown it has a spin liquid ground state with a finite excitation
gap \cite{ED,RH,SPS}.  In the limit of $J' \rightarrow \infty$,
the ground state is an ensemble of singlet rungs and the total spin
is trivially $S=0$.  Low lying excitations are Bloch states of one
triplet with energy $J'+J \cos k_x$, and have a minimum at $k_x = \pi$.
With decreasing $J'/J$, the dispersion becomes more linear with a
smaller gap, but at $J'=J$ the gap remains finite
($\Delta_{\rm spin} = 0.5J$).

The first evidence of enhanced superconducting fluctuations in the
$t$-$J$ ladder is the formation of a bound pair of two holes doped into
half-filling. The ground state wave function is calculated by the
Lanczos method up to $L=10$ and found to have $S=0$ and ${\bf k}=(0,0)$.
The two holes have a positive binding energy defined by,
$  E_{\rm B} \equiv 2E_{\rm GS} (\mbox{2$L$$-$1})
 - E_{\rm GS} (2L)-E_{\rm GS} (\mbox{2$L$$-$2})  $,
where $E_{\rm GS}(N)$ denotes the ground state energy for $N$ electrons.
In Fig.\ \ref{figgap}, we show $E_{\rm B}$ for $L=8$ and $J=0.3$ as a
function of the rung exchange, $J'$. The binding energy is positive
down to the isotropic value, $J'=J=0.3$ appropriate for a cuprate.
The same calculation for four holes indicates no phase separation
up to the largest value $J'=3.0$.

In the limit of strong rung coupling $J' \gg J,1$, this binding can
be easily understood.  At half-filling every rung forms a spin singlet.
A single hole breaks a singlet so the second hole is attracted to the
same rung to avoid breaking another singlet bond.  The binding energy
is therefore $E_B \sim J'-2$ asymptotically.

The hole binding in the ground state is also seen directly in the
hole-hole correlation function,
$\langle n_{a h} (j) n_{a' h} (j') \rangle$
($n_{a h} (j) \equiv 1-n_{a} (j) $: hole density).
We calculated the size of bound hole pair $\xi$ by fitting
$ \langle n_{1 h} (i) n_{2 h} (j) \rangle \sim {\rm const.} \times
  (e^{-|i-j|/\xi} + e^{-(L-|i-j|)/\xi}) $ for
$|i-j|={L \over 2}$ and ${L \over 2}-1$.
The results are shown in the inset of Fig.\ \ref{figgap} for $L=6,8$,
and 10.  In the strong coupling limit $J' \gg J,1$, the two holes are
confined to the same rung and $\xi \rightarrow 0$.  The size of the
bound hole pair increases with decreasing $J'$, but is still quite short,
$\xi \approx 2$, even for isotropic coupling $J' = 0.3$. Note at this
value, the hole-hole correlation is maximum for holes on neighboring
rungs, {\it i.e.}, $\langle n_{1 h} (j) n_{2 h} (j+1) \rangle$.

One of the most interesting properties of the $t$-$J$ ladder is that
there are two distinct types of spin excitations upon doping. Let us
start from the large $J'$ limit, where they are easily distinguished.
The first is the triplet excitation similar to that at half-filling
and arises when a singlet rung away from the hole pair is excited to
a triplet. The triplet propagates along the ladder with the matrix
element ${J \over 2}$, and can also pass through the hole pair.  For
the second type of spin excitation, the presence of the holes is
essential. The bound hole pair in the ground state dissociates into
two separate holons, each of which is now bound to a spinon on the
same rung to form a QP with charge $+|e|$ and spin ${1 \over 2}$
\cite{KT}.  In this sense, these holon-spinon bound pairs are similar
to QP's in conventional Fermi liquids.

The energies and allowed numbers of the two types of spin excitations
are different.  The triplet can be excited only at the rungs without
holes, whereas the second type needs a hole. Therefore the number of
possible excitations is proportional to $(1-\delta)$ and $\delta$,
respectively, where $\delta$ is the hole doping.

The second type of spin excitations have lower energies in the large
$J'$ region  as follows. For both types at least two singlets have to
be broken with energy cost $J'$, but their kinetic energies differ.
In the first type, rungs are occupied by a hole pair and by a triplet.
The hole pair and the triplet move independently along the ladder with
the matrix elements $(J'-{4 \over J'})^{-1}$ and ${J \over 2}$,
respectively. In the second type, two rungs are occupied by only one
electron in a bonding orbital with kinetic energy gain $(-2)$.
These singly occupied rungs move with the matrix element ${1 \over 2}$.
Therefore the total gain of kinetic energy is much larger for the second
type when $J' \gg 1, J$.  Accordingly the lowest spin triplet excitation
is of the second type as shown in Fig.\ \ref{figgap}, and
$\Delta_{\rm spin} \sim E_{\rm B}$. With decreasing $J'$, the two
excitations become closer to each other in energy and they are strongly
mixed since they can have the same symmetry.

The different characters of the two types of excitations are clearly
distinguished by the hole-hole and spin-hole correlation functions.
Figure \ref{figcor} shows the correlation functions calculated by the
Lanczos method for the two excitations with $S^z =1$ for $J=0.3$ and
$J'=0.3$, and $3.0$.  In the QP excitation (dotted line), the hole-hole
correlation is maximum for the longest separation, but for the spin-hole
correlations on the same rung.  This tendency is obvious for the large
rung coupling $J'=3.0$, but still holds down to the isotropic value
$J'=0.3$.  For the other excitations (dashed line), at $J'=3.0$ the two
holes are strongly bound on the same rung and repel the triplet.
With decreasing $J'$, the hole pair becomes more extended as in the
singlet ground state.  The most dominant configuration at $J'=0.3$ is
that the two holes are in the neighboring rungs and different legs.

The two types of excitations have different contributions to the spin
susceptibility and structure factor,
\begin{equation}
  {\rm Im} \ {\cal S} ({\bf k},\omega)
  \equiv \sum_{\alpha}
  \left| \langle \alpha| S_{\bf k}^+ | {\rm GS} \rangle
  \right| ^2
  \delta ( \omega - E_\alpha + E_{\rm GS} ) .
\label{eqstf}
\end{equation}
Here $|\alpha\rangle$ and $|{\rm GS}\rangle$ denote an $S=1$ eigenstate
and the ground state with energies $E_\alpha$ and $E_{\rm GS}$.
As was discussed before the spin excitations with separate holes have
lower energies.  However at $\delta \ll 1$, their number is smaller
($\propto \delta$) than the excitations with the bound hole pairs and
triplet rungs ($\propto 1-\delta$).  Therefore with decreasing
temperature the susceptibility will show a large exponential drop at
$T \sim J'$ and a smaller drop further at $T \sim \Delta_{\rm spin}$,
corresponding to the different excitation energies.
Figure \ref{figstf} shows the structure factor,
$ {\rm Im} \ {\cal S} ({\bf k},\omega) $, for two holes in
$8 \times 2$ sites.  Large peaks are seen around ${\bf k}=(\pi,\pi)$
and the energy of the order of $J'$.  These are due to the excitations
away from bound hole pairs.  On the other hand, the QP excitations do
not have large weights in spite of their lower excitation energies,
because of the change in the charge configuration from the ground
state.  This point would be important when comparing to neutron
scattering experiments.

Finally we discuss the one-particle Green's function
where we can see QP excitations directly.
By using the Lanczos method combined with a continued fraction method,
we calculate the spectral function for two holes.
The electron and hole parts of the spectral function are defined as
\begin{eqnarray}
  A_{\rm e} ({\bf k},\omega)
  && \equiv \sum_\alpha
    | \langle \alpha, \mbox{2$L$$-$1}
           | c_{{\bf k}\sigma}^\dagger
           |{\rm GS},\mbox{2$L$$-$2}
      \rangle
    |^2
  \nonumber\\
  && \times
     \delta (
              \omega - E_\alpha   (\mbox{2$L$$-$1})
                     + E_{\rm GS} (\mbox{2$L$$-$2}) + \mu
            ),
  \nonumber\\
  A_{\rm h} ({\bf k},\omega)
  && \equiv \sum_\alpha
    | \langle \alpha, \mbox{2$L$$-$3}
           | c_{{\bf k}\sigma}
           |{\rm GS},\mbox{2$L$$-$2}
      \rangle
    |^2
  \nonumber\\
  && \times
     \delta (
              \omega + E_\alpha   (\mbox{2$L$$-$3})
                     - E_{\rm GS} (\mbox{2$L$$-$2}) + \mu
            ),
\label{eqspe}
\end{eqnarray}
where $|\alpha , N \rangle$ is an eigenstate for $N$ electrons
with the energy $E_\alpha (N)$ and GS denotes the ground state.
Positive (negative) energies correspond to the electron (hole) part.
The chemical potential is defined by
$ \mu \equiv {1 \over 2}
  [ E_{\rm GS}(\mbox{2$L$$-$1})-E_{\rm GS}(\mbox{2$L$$-$3}) ]$.
The results are shown in Fig.\ \ref{figspe} for $L=8$ and
$J = J' = 0.3$. The wavevector along the ladder
$k_x = {2 \pi \over L} n$ ($n$: integer) are for PBC and
$k_x = {2 \pi \over L} (n + {1 \over 2})$ for APBC.  The ground state
energy $E_{\rm GS}(2L-2)$ and the chemical potential $\mu$ in
Eq.\ (\ref{eqspe}) are the average over both boundary conditions.

There are large weights for the bonding ({\sl B})($k_y =0$) and
antibonding ({\sl A})($k_y = \pi$) orbitals only near the Fermi energy
$\omega = 0$, and they seem to constitute QP bands.  Away from the
Fermi energy, the individual QP peaks are much less prominent and
there is an incoherent part with the energy of the order of $1$.

The QP spectrum is consistent with the MF theory based on the
$d$-wave RVB state. Without holes the {\sl B}- and {\sl A}-bands
are degenerate because the kinetic energy is zero. Upon hole doping,
$\langle c_{1 \sigma}^\dagger (j) c_{2 \sigma} (j) \rangle$
on a rung becomes finite and the two bands are split.
The {\sl B}-band is pushed down and occupied by more electrons,
while the {\sl A}-band is pushed up and occupied by less electrons.
The QP with energy closest to $\omega =0$ has a wavevector nearest
to the original Fermi $k_F$: ($k_x = {5\pi \over 8}$ for {\sl B} and
$k_x = {3\pi \over 8}$ for {\sl A}). Because of the band splitting,
$k_F^{\rm B} > k_F^{\rm A}$, but the Luttinger sum rule is satisfied,
$k_F^{\rm B} + k_F^{\rm A} = (1-\delta) \pi $.
This means the Fermi surface is large, consistent with
photoemission experiments on cuprates.

It is important to notice the QP peaks near the Fermi energy have
their counterparts on the opposite side of the Fermi energy.  An
electronic QP peak at energy $\omega >0$ has a shadow hole peak at
energy around $-\omega <0$, and {\it vice versa}.  These peaks
indicate that the QP excitations are those of the Bogoliubov QP's
as in BCS theory, {\it i.e.}, mixture of an electron and a hole
($ \alpha_{\bf k}^\dagger =
   u_{\bf k} c_{{\bf k} \uparrow}^\dagger +
   v_{\bf k} c_{-{\bf k} \downarrow}$).
The weights in the electron and hole parts are proportional to
$|u_{\bf k}|^2$ and $|v_{\bf k}|^2$. They are hole-like around
$k_x =0$ and electron-like around $k_x =\pi$ for both {\sl B} and
{\sl A} bands. There exists a finite energy gap in the QP spectra.
The electron and hole branches both come close to the Fermi energy
at $k_x \sim {\pi \over 2}$, but instead of passing through they move
away from it.  The energy gap is $0.18t$ at
${\bf k}=({5\pi \over 8},0)$ and $0.23t$ at
${\bf k}=({3\pi \over 8},\pi)$, corresponding to a QP gap
$2\Delta_{\rm QP} \simeq 0.2 (\simeq {2J \over 3})$.
It is interesting to note that the Stephan-Horsch results for
$A({\bf k},\omega)$ in two-dimensional clusters \cite{WS}
show similar behavior for ${\bf k}$-points not along $(1,1)$
but no shadow peaks for ${\bf k} \parallel (1,1)$,
indicating $d_{x^2-y^2}$-pairing also.

The results of our exact diagonalization studies show that
the doped $t$-$J$ ladder belongs to a different universality
class than the single chain.  Here the separation is into a
limited number of quasiparticle excitations carrying charge
$+|e|$ and spin ${1 \over 2}$ which vanish as $\delta \rightarrow 0$,
and a triplet collective mode which remains as $\delta \rightarrow 0$.
So the triplet mode is {\em not} a collective mode of the
quasiparticles.  This separation resembles in certain ways the
proposal of Sokol and Pines for a doped quantum critical
regime \cite{SP} and also Chubukov and Sachdev for collective
and Fermion contributions to the susceptibility \cite{CS}.
Note the quasiparticles do not behave as a usual Fermi liquid.
Although their number is strictly limited when $\delta \ll 1$,
yet their pairing energy is finite as $\delta \rightarrow 0$.
In this sense they resemble a dilute Fermi gas with a strong
attraction giving pair binding \cite{R}.  However if we look
at the dispersion energy of an added hole that would be seen in
photoemission, then from Fig.\ \ref{figspe} it is characterized
by the large Fermi surface.  The quasiparticles do not form a
usual Fermi liquid but display a new and interesting mixture of
dilute Fermi gas and large Fermi surface behavior.

The authors thank S. Gopalan, H. Monien, D. Poilblanc, D. W\"urtz,
F. C. Zhang, and especially M. Sigrist for fruitful discussions.
The work was supported by the Swiss National Science Foundation
Grant No. NFP-304030-032833 and SNF-21-27894.89
and by an internal grant of ETH-Z\"urich.

%
%

%
%


%
%
\begin{figure}
\caption{
Binding energy of two holes, spin gap, and energy of the triplet
excitation away from the bound hole pair. Inset shows the size
of the bound hole pair for the ground state
}
\label{figgap}
\end{figure}
%
%
\begin{figure}
\caption{
Hole-hole ($(a)$,$(c)$) and spin-hole ($(b)$,$(d)$) correlation
functions for the two triplet states.
Dotted lines are for the lowest triplet state (QP excitation)
and dashed lines are for the triplet state with a bound pair.
}
\label{figcor}
\end{figure}
%
%
\begin{figure}
\caption{
Spin structure factor, ${\rm Im} \ {\cal S} ({\bf k},\omega)$,
for the $t$-$J$ ladder with two holes and APBC.
$L=8$ and $J = J' = 0.3$.
The width of each line represents the strength of the excitation.
}
\label{figstf}
\end{figure}
%
%
\begin{figure}
\caption{
Spectral function of the one-particle Green's function,
$A({\bf k},\omega)$, for two holes. $L=8$ and $J = J' = 0.3$.
The width of each line represents the strength of the excitation.
}
\label{figspe}
\end{figure}


\begin{references}

\bibitem{ED}
E. Dagotto, J. Riera, and D. J. Scalapino, \prb {\bf 45}, 5744 (1992);
T. Barnes et al., \prb {\bf 47}, 3196 (1993).

\bibitem{RGS}
T. M. Rice, S. Gopalan, and M. Sigrist, Europhys. Lett.
{\bf 23}, 445 (1993).

\bibitem{MS}
M. Sigrist, T. M. Rice, and F. C. Zhang, preprint, ETH-TH/93-35.

\bibitem{RMN}
R. M. Noack, S. R. White, and D. J. Scalapino, preprint.

\bibitem{SG}
S. Gopalan, T. M. Rice, and M. Sigrist, preprint, ETH-TH/93-39.

\bibitem{DVK}
D. V. Khveshchenko and T. M. Rice, preprint;
D. V. Khveshchenko, preprint.

\bibitem{RH}
R. Hirsch, Diplomarbeit, University of K\"oln (1988).

\bibitem{SPS}
S. P. Strong and A. J. Millis, \prl {\bf 69}, 2419 (1992).

\bibitem{MO}
M. Ogata, M. U. Luchini, and T. M. Rice, \prb {\bf 44}, 12083 (1991).

\bibitem{MI}
M. Imada, \prb {\bf 48}, 550 (1993), and references therein.

\bibitem{MT}
M. Takano et al., JJAP Series {\bf 7}, 3 (1992).

\bibitem{DCJ}
D. C. Johnston et al., \prb {\bf 35}, 219 (1987).

\bibitem{KT}
Similar excitations are also found for a $t$-$J$ chain
with frustrated couplings.
K. Takano and K. Sano, \prb {\bf 48}, 9831 (1993);
I. Bose and S. Gayen, \prb {\bf 48} 10653 (1993).

\bibitem{WS}
W. Stephan and P. Horsch, \prl {\bf 66}, 2258 (1991).

\bibitem{SP}
A. Sokol and D. Pines, \prl {\bf 71}, 2813 (1993).

\bibitem{CS}
A. V. Chubukov and S. Sachdev, \prl {\bf 71}, 169 (1993).

\bibitem{R}
C. A. R. S\'a de Melo, M. Randeria, and J. R. Engelbrecht,
\prl {\bf 71}, 3202 (1993).

\end{references}
\end{document}